\documentclass[final,english]{bullsrsl}[2022/06/15]
\usepackage[latin1]{inputenc}
\usepackage[T1]{fontenc}
\usepackage{natbib} 
\usepackage{graphicx}
\usepackage{amsmath}

\begin{document}
\title{Study of the Balmer decrements for Galactic classical Be stars using the Himalayan Chandra Telescope of India}

\author[affil={1}, corresponding]{Gourav}{Banerjee}
\author[affil={1}]{Blesson}{Mathew}
\author[affil={1}]{Suman}{Bhattacharyya}
\author[affil={1}]{Ashish}{Devaraj}
\author[affil={1}]{Sreeja}{Kartha}
\author[affil={2}]{Santosh}{Joshi}

\affiliation[1]{Department of Physics and Electronics, CHRIST (Deemed to be University), Hosur Main Road, Bangalore, India}
\affiliation[2]{Aryabhatta Research Institute of Observational Sciences, Nanital, India}

\correspondance{gourav.banerjee@res.christuniversity.in}
\date{13th October 2020}
\maketitle

\begin{abstract}
In a recent study, \cite{2021Banerjee} produced an atlas of all major emission lines found in a large sample of 115 Galactic field Be stars using the 2-m Himalayan Chandra Telescope (HCT) facility located at Ladakh, India. This paper presents our further exploration of these stars to estimate the electron density in their discs. Our study using Balmer decrement values indicate that their discs are generally optically thick in nature with electron density ($n_e$) in their circumstellar envelopes (CEs) being in excess of 10\textsuperscript{13}\,cm\textsuperscript{-3} for around 65\% of the stars. For another 19\% stars, the average $n_e$ in their discs probably range between 10\textsuperscript{12}\,cm\textsuperscript{-3} and 10\textsuperscript{13}\,cm\textsuperscript{-3}. We noticed that the nature of the H$\alpha$ and H$\beta$ line profiles might not influence the observed Balmer decrement values (i.e. $D_{34}$ and $D_{54}$) of the sample of stars. Interestingly, we also found that around 50\% of the Be stars displaying $D_{34}$ greater than 2.7 are of earlier spectral types, i.e. within B0 -B3.
\end{abstract}

\keywords{Be star, spectroscopy, emission lines, Balmer decrement, variability}


\section{Introduction}
A classical Be (Be hereafter) star is a B-type, massive main sequence star surrounded by a circumstellar equatorial, gaseous decretion disc which is geometrically thin in nature and orbits the central star in Keplerian rotation \citep{2006Carciofi, 2007Meilland}. Spectra of Be stars show emission lines of different elements. Spectroscopic line profile variability is another commonly observed property of Be stars \cite[e.g.][]{2022Banerjee, 2017Paul, 2003Porter, 2002Miroshnichenko}. Studying these lines and their variability provide an excellent opportunity to understand the geometry and kinematics of the circumstellar disc and properties of the central star itself. However, the disc formation mechanism in Be stars -- the `Be phenomenon' -- is still poorly understood. As a result, spectroscopic study of Be stars has gained momentum in the respective community within the past decades to better understand the `Be phenomenon' using different national optical facilities.

For example, \cite{2008Mathew} performed a slitless spectroscopic survey to study the spectral features of 150 Be stars in open clusters. The spectral features seen in these stars were presented in \cite{2011Mathew}. Following this, \cite{2021Banerjee} performed a spectroscopic study of all major emission lines for a sample of 115 field Be stars in the Galaxy in the wavelength range of 3800 - 9000~\AA~using the 2-m Himalayan Chandra Telescope (HCT) facility located at the Indian Astronomical Observatory (IAO), Ladakh, India. Considering the potential of the obtained data, in this paper, we have further explored these stars to better understand their disc properties.

Moreover, quite a number of recent works by different authors have identified and studied various types of emission-line stars \cite[e.g.][]{2023Jagadeesh, 2022Bhattacharyya, 2022Shridharan, 2021Bhattacharyya, 2021Jagadeesh, 2021Wang, 2021Li, 2021Anusha}. Detection and study of more Be stars in fields and both younger and older clusters may provide new insights about the `Be phenomenon' in diverse environments. Towards this objective, we became motivated to perform some further analysis of the data of Be stars obtained by the HCT facility.

\section{Observations}
The spectroscopic observations of the 115 Be stars were carried out with the Himalaya Faint Object Spectrograph Camera (HFOSC) instrument mounted on the 2-m HCT in Ladakh at the Cassegrain focus. This instrument is equipped with a 2K$\times$4K SiTe CCD system with every pixel corresponding to 0.3$\times$0.3\,arcsec\textsuperscript{2}. This covers an area of 10$\times$10\,arcmin\textsuperscript{2}. The gain and readout noise for the HFOSC instrument are 1.22\,e$^{-}$\,ADU$^{-1}$ and 4.8\,e$^{-}$, respectively.

During December 2007 to January 2009 we observed 115 Be stars selected from the catalogue of \cite{1982Jaschek}. These stars were selected based on the observation visibility of HCT. The spectral coverage is from 3800 -- 9000 \AA. The spectrum in the `blue region' is taken with Grism 7 (3800 $-$ 5500 \AA), which in combination with 167l slit provides an effective resolution of 10 \AA~at H$\beta$. The red region spectrum is taken with Grism 8 (5500 $-$ 9000 \AA) and 167l slit, providing an effective resolution of 7 \AA~at H$\alpha$. Dome flats taken with halogen lamps were used for flat fielding the images. Bias subtraction, flat field correction and spectral extraction were performed with standard IRAF tasks. FeNe and FeAr lamp spectra were taken with the object spectra for wavelength calibration. All the extracted raw spectra were wavelength calibrated and continuum normalized with IRAF tasks.

\section{Discussion on previous results}
Our previous study \citep{2021Banerjee} provided valuable insights about the nature of Be star discs. We made use of the unprecedented capability of the {\it Gaia} mission to re-estimate the extinction parameter (A$_V$) for these stars. The analysis of spectral lines considering the {\it Gaia} DR2 data added more meaning to our study. As an important outcome, we were able to use the estimated A$_V$ values for extinction correction in the analysis of the Balmer decrement ($D_{34}$ and $D_{54}$) values for the program stars. In Be stars, relative emission strengths of Balmer emission lines is a function of the electron temperature and density in their disc. The flux ratios for the strongest emission lines visible in Be star optical spectra are known as Balmer decrements. Thus, Balmer decrement is defined as

\begin{equation}
        D_{34} = F(H_\alpha) / F(H_\beta)\\
\end{equation}

\begin{equation}
        D_{54} = F(H_\gamma) / F(H_\beta)\\
\end{equation}

Here, $D_{34}$ and $D_{54}$ are usually quantified with respect to the emission strength (i.e. flux) of the H$_\beta$ line, F(H$_\beta$). This is primarily used for better understanding of the electron temperature and density in Be star discs.

We estimated the $D_{34}$ values for 105 and $D_{54}$ values for 96 of the program stars following the method described in \cite{2021Banerjee}. We found the $D_{34}$ value in our sample to range between 0.1 and 9.0, whereas the corresponding $D_{54}$ value mostly ($\approx$ 70\%) ranges between 0.2 and 1.5, clustering somewhere near 0.8 $-$ 1.0. Our study suggested that Be star discs are generally optically thick in nature in majority of the cases. This study also highlighted the importance of considering A$_V$ for studying Be star properties.

\section{Results}

\subsection{Evaluation of the electron density in Be star discs using Balmer decrement values}
\begin{figure}[t]
\centering
\includegraphics[width=\textwidth]{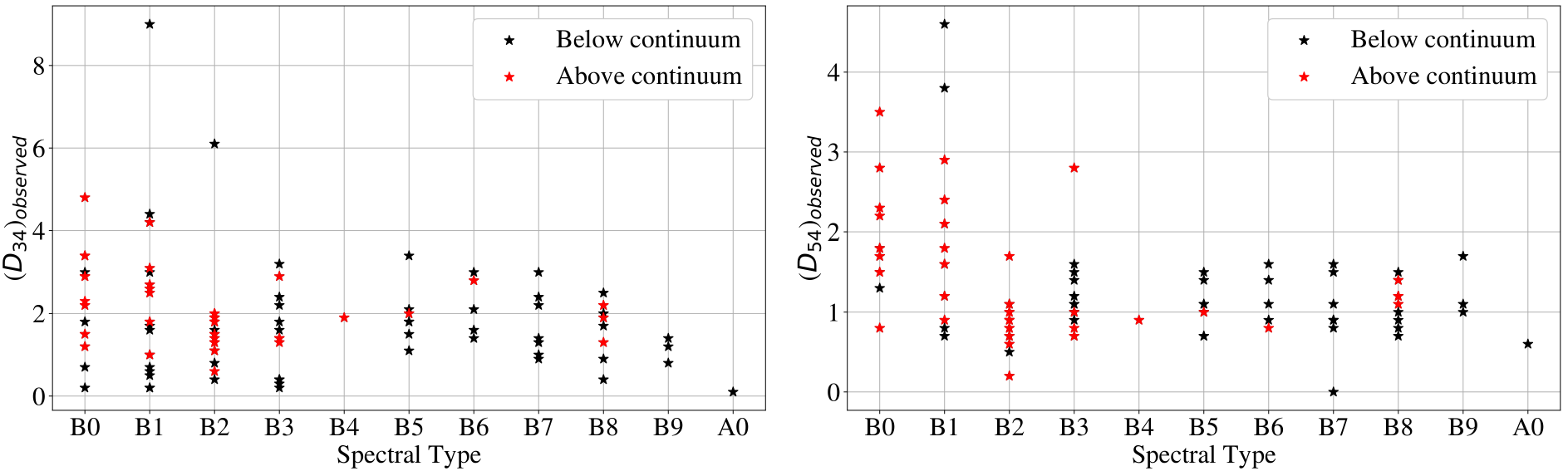}
\begin{minipage}{12cm}
\caption{Spectral type distribution of 105 and 96 stars against their observed $D_{34}$ and $D_{54}$ values. Stars exhibiting both H$\alpha$ and H$\beta$ lines above the continuum are marked with red star symbols, whereas the other stars are marked with black star symbols.}
\label{D34val}
\end{minipage}
\end{figure}

Before evaluating the electron density in the discs of the sample of Be stars, we wanted to verify whether the nature of line profiles can influence the observed $D_{34}$ and $D_{54}$ values or not. Spectral type distribution of 105 and 96 stars against their observed $D_{34}$ and $D_{54}$ values is shown in Fig.\,\ref{D34val}. We found that 39 among 105 stars show both H$\alpha$ and H$\beta$ lines above the continuum, i.e. clear emission. However, for rest of the stars, emission in either H$\alpha$ or H$\beta$ or for both H$\alpha$ and H$\beta$ lines are found to exist below the continuum. Since in-filling of photospheric absorption lines occur, only visual inspection cannot always discern between a star showing weak emission and another one exhibiting no emission at all. So to calculate the H$\alpha$ and H$\beta$ equivalent width (EW) we at first measured the EW values for all these stars. Then we measured the absorption component at H$\alpha$ and H$\beta$ lines from synthetic spectra using models of stellar atmospheres \citep{1993Kurucz} for each spectral type. The photospheric contribution from the underlying star is added to the emission component to estimate the corrected H$\alpha$ and H$\beta$ EW \citep{2021Banerjee}. This process of inspection is useful to identify whether H$\alpha$ or H$\beta$ emission actually exists in case of any Be star, even if any or both these lines might appear in absorption in the spectrum.

In Fig.\,\ref{D34val}, those stars showing both H$\alpha$ and H$\beta$ lines above the continuum are marked with red star symbols. The spectra for rest of the stars exhibit either or both H$\alpha$ and H$\beta$ lines in absorption. These stars are represented by black star symbols in the figure. Some of these stars do have hidden emission that exists below the continuum \citep{2021Banerjee}. The EW of H$\alpha$ and H$\beta$ lines might be over-estimated in some cases. So we considered the measured EW of H$\alpha$ and H$\beta$ lines for making this plot.

From the figure, it is observed that no distinct trend can be detected. Although in the left panel plot we can see 5 points within B0 -- B2 showing ($D_{34})_{observed}$ > 4, any particular trend is difficult to detect.  Interestingly, the overall value of ($D_{54})_{observed}$ appears to decrease for later spectral types. It will be possible to confirm whether this is a real trend or not through further studies using larger sample size. However, the figure points out that the nature of the H$\alpha$ and H$\beta$ line profiles might not influence the ($D_{34})_{observed}$ and ($D_{54})_{observed}$ values of the sample of Be stars. Next, we performed a comparative study with the existing literature to evaluate the electron density in their discs using the $D_{34}$ and $D_{54}$ values.

It is presently well established that disc models for Be star circumstellar envelopes (CEs) are consistent with observed data if the typical electron temperatures ($T_e$) in the discs are 10\textsuperscript{4} K, electron densities ($n_e$) are of the order of around 10\textsuperscript{12}\,cm\textsuperscript{-3}, and when the disc radii ($R_d$) range within 10\textsuperscript{12} and 10\textsuperscript{13}\,cm \citep{1990Dachs}. However, for individual Be stars there exist obvious differences between properties of CEs. Also, variability of CEs observed in many Be stars needs to be explained in terms of variations and differences of their different physical parameters, such as dimensions and electron densities of the CEs.

In this regard, it becomes necessary to check if, or how our estimated Balmer decrements may serve as a possible tool of supplementary information to shed light on the optical thickness prevailing in Be star CEs. To gain insights about possible electron densities occurring in Be star CEs it is important to perform a comparative study of our estimated Balmer decrement values with the corresponding values measured for different kinds of model gaseous nebulae by previous authors. In an important study, \cite{1971Brocklehurst} reported that for a gaseous nebula under case B conditions, having $T_e$ = 1 $\times$ 10\textsuperscript{4} K, $n_e$ $\leq$ 10\textsuperscript{6}\,cm\textsuperscript{-3}, the expected $D_{34}$ and $D_{54}$ values can be 2.85 and 0.47, respectively.

\cite{1980Drake} computed the theoretical Balmer decrements considering a static, high-density nebula having slab geometry. Such type of model nebulae were devised in order to model line emission from Active Galactic Nuclei (AGNs) and quasars \citep{1990Dachs}. For their calculations, \cite{1980Drake} adopted a model hydrogen atom containing 20 bound levels and covered particle density range from 10\textsuperscript{8} to 10\textsuperscript{15}\,cm\textsuperscript{-3}. In case of the strongest radiation fields and considering $T_e$ = 10\textsuperscript{4} K, ground state photoionization rate, $R_1c$ = 3\,s\textsuperscript{-1} and $\tau$(Ly$\alpha$)  = 10\textsuperscript{5}, \cite{1980Drake} (Fig.\,12c) found $D_{34}$ to continuously decrease as electron density ($n_e$) increases from a maximum value of around 5 at $n_e$ $\sim$ 10\textsuperscript{10}\,cm\textsuperscript{-3} to 1 at $n_e$ $\sim$ 10\textsuperscript{15}\,cm\textsuperscript{-3}. The corresponding $D_{54}$ value was noticed to increase accordingly from 0.34 to 0.65 (Fig.\,13g). Considering the same conditions these authors found $D_{34}$ = 3.3 and $D_{54}$ = 0.46 at $n_e$ $\sim$ 10\textsuperscript{12}\,cm\textsuperscript{-3}. 

In a separate work, studying model hydrogen nebulae at $T_e$ = 10\textsuperscript{4} K and for a column density ($N_H$) = 10\textsuperscript{23} hydrogen atoms cm\textsuperscript{-2}, \cite{1987Joly} obtained slightly different results. The author found $D_{34}$ to continuously decrease from $D_{34}$ = 3.3 at $n_e$ = 10\textsuperscript{10}\,cm\textsuperscript{-3} to $D_{34}$ = 2.7 at $n_e$ = 10\textsuperscript{12}\,cm\textsuperscript{-3}. Another major study by \cite{1988Williams} considered rotating high-density accretion disc models and computed the Balmer decrements for line emission originating from such discs. However, we did not consider their results for the present study since Be star discs are decretion discs, having no dust in their discs. \cite{1990Dachs} performed a comparative study between the results obtained by \cite{1980Drake} and \cite{1987Joly}, which suggested that in the range of $n_e$ = 10\textsuperscript{10}\,cm\textsuperscript{-3} and 10\textsuperscript{12}\,cm\textsuperscript{-3}, Balmer decrement values tend to come closer to the conventional case B values when photoionization from excited levels are included into the model calculations.

From our study, we observed that for 63 ($\sim$ 65\%) out of 96 stars, flat Balmer decrements are observed with $D_{34}$ $\leq$ 2.0, $D_{54}$ $\geq$ 0.7. Only five stars, namely HD 33461, HD 37967, HD 45910, HD 61205 and HD 65079 are the exceptions where it is found that $D_{34}$ $\leq$ 2.0 and also $D_{54}$ $\leq$ 0.7. Most of these 63 stars also show relatively faint H$\alpha$ emission with corrected H$\alpha$ EW < -25 \AA. According to calculations for different high-density model hydrogen nebulae mentioned above, such flat Balmer decrements (i.e. $D_{34}$ $\leq$ 2.0 and also $D_{54}$ $\geq$ 0.7) indicate the presence of higher electron densities in the CEs of Be stars as compared to those stars showing H$\alpha$ EW greater than -25 \AA. According to \cite{1980Drake} (Fig.\,12c), in such discs with flat Balmer decrements, $n_e$ will be $\geq$ 10\textsuperscript{13}\,cm\textsuperscript{-3}. In similar conditions, \cite{1978Krolik} (Table\,4; models DX2, DX3) predicted the $n_e$ to be at least $\geq$ 10\textsuperscript{12}\,cm\textsuperscript{-3}.

We found that the best agreement between our estimated $D_{34}$ and $D_{54}$ values for these 63 stars and theoretically calculated values are obtained if $n_e$ in the CEs of those stars are considered to be in excess of around 10\textsuperscript{13}\,cm\textsuperscript{-3} for the most likely gaseous nebulae models calculated by \cite{1980Drake}. For another 18 ($\sim$ 19\%) stars where strong H$\alpha$ emission is noticed with EW greater than -25 \AA, average $n_e$ in their CEs probably range between 10\textsuperscript{12}\,cm\textsuperscript{-3} and 10\textsuperscript{13}\,cm\textsuperscript{-3}. We can suggest this by comparing our obtained values with those obtained by theoretical calculations by authors such as \cite{1980Drake} and \cite{1978Krolik} for their model nebulae. Hence, our study suggests that the discs of our sample of Be stars are generally optically thick in nature with electron density ($n_e$) in their CEs being in excess of 10\textsuperscript{12}\,cm\textsuperscript{-3} in majority of the cases. Interestingly, we also found 19 stars which show $D_{34}$ value greater than 2.7. It has already been mentioned that theoretical Balmer decrements calculated for model nebulae by several authors predict that $D_{34}$ value increases with decreasing values of $n_e$. So these 19 stars are interesting cases and need further investigation.

However, it is important to note that the estimated Balmer decrements for Be stars represent the results of averaging over different portions of their CEs exhibiting a definite range of (local) $D_{34}$ and $D_{54}$ values. Hence, it is necessary to develop better modelling of theoretical Balmer decrements for hydrogen line emission originating from Be star CEs in future to obtain a more precise interpretation of estimated $D_{34}$ and $D_{54}$ values.

\subsection{Be stars showing $D_{34}$ value greater than 2.7}
For the 19 stars showing $D_{34}$ value greater than 2.7 it is understood that the H$\alpha$ emission strength is more than H$\beta$, which contributes to larger values of $D_{34}$. Our study points out that the mean electron density (i.e. $n_e$) in their discs are lesser making their discs optically thin in nature.

Hence, we looked into the literature to check the nature of these 19 Be stars. We found that 15 ($\sim$ 79\%) among the 19 stars are reported to be of earlier spectral types (within B0 -B3) in SIMBAD.  Another star, HD 259431 has been identified to be a Herbig Ae/Be star \citep{1998Waters}. The rest three stars, namely HD 72043, HD 237118 and HD 37115 have spectral types B5, B6 and B7, respectively. HD 72043 is a less studied object, identified as a Be star by \cite{1982Jaschek}. Interestingly, the luminosity classes of 13 out of these 18 stars are reported to be V indicating them to be in the main sequence (MS) phase. Among the rest 86 stars which display $D_{34}$ $\leq$ 2.7, 56 belong within B0 and B3  spectral types.

Looking into the literature, it is also found that \cite{2021Bhattacharyya} classified two among the 19 stars, namely CD-22 4761 and BD-11 2043 as Transition phase (TP) candidates, which are rare emission-line stars in transition phase from pre-main sequence (PMS) to main sequence. It is expected that such stars will possess dust component in their discs. One other star, HD 55606 has been identified as a Be+sdO binary system by \cite{2018Chojnowski}. Another star, HD 50820 is reported to be a binary star with composite K4III++B1.5Ve spectral type \citep{2002Ginestet}.

We then checked \cite{1990Dachs} who reported 6 among their sample of 26 Be stars (HR 2787, HR 2911, HR 4140, HR 6510, HR 8402 and HR 8773) to show $D_{34}$ greater than 2.7 on every epoch of their observations. They also found the $D_{54}$ value for 5 among these 6 stars to range between 0.36 and 0.5 on all dates of observations. Checking the literature, we found that only one among these 6 stars (HR 2787) is reported to be a binary. The spectral types for these 6 stars range within B2 and B6, with three (50\%) of them (HR 2787, HR 2911 and HR 6510) belonging to B2 type. So from our present study it is interesting to note that around 50\% of the Be stars displaying $D_{34}$ greater than 2.7 are of earlier spectral types, i.e. within B0 -- B3.

\section{Conclusions}

\begin{enumerate}


\item Our results indicate that H$\alpha$ and H$\beta$ line profiles do not have any influence on the observed $D_{34}$ and $D_{54}$ values of the sample stars.

\item In addition, our initial study suggests that Be star discs are generally optically thick in nature with electron density ($n_e$) in their circumstellar envelopes (CEs) being in excess of 10\textsuperscript{12}\,cm\textsuperscript{-3} in majority of the cases. We suggest that $n_e$ in the CEs of 63 (having corrected H$\alpha$ EW lesser than -25 \AA) out of 96 stars are possibly in excess of around 10\textsuperscript{13}\,cm\textsuperscript{-3}, whereas for another 41 stars (showing corrected H$\alpha$ EW greater than -25 \AA) average $n_e$ in their discs probably range between 10\textsuperscript{12}\,cm\textsuperscript{-3} and 10\textsuperscript{13}\,cm\textsuperscript{-3}.

\item Interestingly, it is noticed that around 50\% of the Be stars displaying $D_{34}$ greater than 2.7 are of earlier spectral types, i.e. within B0 -- B3.
\end{enumerate}

\section{Acknowledgement}
We would like to thank the staff of the Indian Astronomical Observatory (IAO), Ladakh for taking the observations using the HCT facility situated at Hanle. We also thank the Center for research, CHRIST (Deemed to be University), Bangalore, India. BM acknowledges the support of the Science \& Engineering Research Board (SERB), a statutory body of the Department of Science \& Technology (DST), Government of India, for funding our research under grant number CRG/2019/005380. Also, SB is grateful to the Centre for Research, CHRIST (Deemed to be University), Bangalore for the research grant extended to carry out the present project (MRP DSC-1932). Moreover, this work has used the {\it Gaia} DR2 data to re-estimate the extinction parameters for the program stars. Hence, we express our gratitude to the {\it Gaia} collaboration for providing the data. Finally, both GB and SB would like to acknowledge the local support provided by the BINA project during the time of BINA-2023 workshop.

\begin{furtherinformation}

\begin{orcids}
\orcid{0000-0001-8873-1171}{Gourav}{Banerjee}
\orcid{0000-0002-7254-191X}{Blesson}{Mathew}
\orcid{0000-0002-1920-6055}{Suman}{Bhattacharyya}
\orcid{0000-0001-5933-058X}{Ashish}{Devaraj}
\orcid{0000-0002-7666-1062}{Sreeja}{Kartha}
\orcid{0009-0007-1545-854X}{Santosh}{Joshi}
\end{orcids}

\begin{authorcontributions}
This work is part of a collective effort where all co-authors provide contributions.
\end{authorcontributions}

\begin{conflictsofinterest}
The authors declare no conflict of interest.
\end{conflictsofinterest}

\end{furtherinformation}






\bibliographystyle{bullsrsl-en}

\bibliography{S06-CT03_BanerjeeG}

\end{document}